# Blockchain Applications in Power Systems: A Bibliometric Analysis

Hossein Mohammadi Rouzbahani, Hadis Karimipour , Ali Dehghantanha, Reza M. Parizi

University of Guelph, hmoham15@uoguelph.ca

University of Guelph, hkarimi@uoguelph.ca

University of Guelph, adehghan@uoguelph.ca

Kennesaw State University, rparizi1@kennesaw.edu

**Abstract.** Power systems are growing rapidly, due to ever-increasing demand for electrical power. These systems require novel methodologies and modern tools and technologies, to better perform, particularly for communication among different parts. Therefore, power systems are facing new challenges such as energy trading and marketing and cyber threats. Using blockchain in power systems, as a solution, is one of the newest methods. Most studies aim to investigate innovative approaches of blockchain application in power systems. Even though, many articles published to support the research activities, there has not been any bibliometric analysis which specifies the research trends. This paper aims to present a bibliographic analysis of the blockchain application in power systems related literature, in the Web of Science (WoS) database between January 2009 and July 2019 . This paper discusses the research activities and performed a detailed analysis by looking at the number of articles published, citations, institutions, research area, and authors. From the analysis, it was concluded that there are several significant impacts of research activities in China and USA, in comparison to other countries.

## 1. Introduction

Power systems are experiencing swift changes due to the rapid growth of electricity demand, which is expected to grow further by around 30% in 2035 [1], [2]. While, power systems are facing challenges because of new technology developments, security concerns, new market patterns, consumer demand changes, etc.

[3], [4]. There are some permanent challenges such as stability, reliability, environmental concerns, and costs [5].

Various type of methods have been employed over the years for solving problems in different sections and improving the performance of the power systems [3], [6], [7]–[15], but some of these problems are related to the growing network and its integration. Due to using distributed generation in power systems (which is one of the main reasons for network growth) and using communication tools and smart meters, marketing and communication in power networks require new and up-to-date methods. It should be noted power systems are on the edge of entering the digital era by a massive deployment of in most countries in the world [3], [16]. Figure.1 shows the summary of the content discussed.

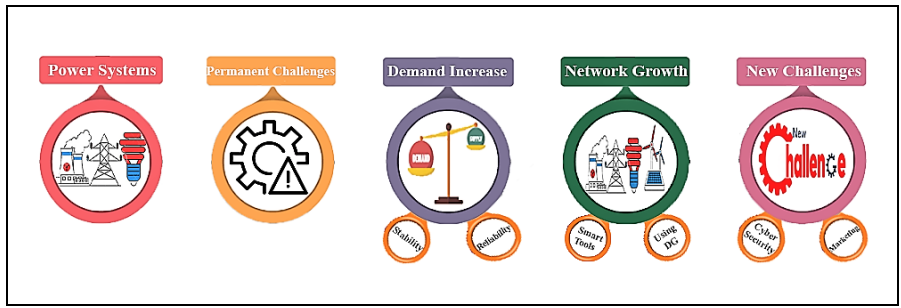

**Fig. 1 Power system changes and challenges**

Central management and operation are becoming ever more challenging because of the need for an advanced communication and data exchanges among different parts of the power network [17]. On the other hand, the decentralization of the property, and the decision-making process are complex which are evolving the Information Technologies [18], [19]. Thus, to accommodate these decentralization and digitalization trends, local distributed control and management techniques are in need [20]. The importance of this issue has led many researchers to seek new methodologies and concepts to improve the performance and security of the power systems [8], [13], [21], [22]. Application of blockchain is one of the newest ones.

Blockchain, also called distributed ledger, is a technology, by a set of nodes that do not fully trust each other, which first was proposed in 2008 [23], [24] . This technology is designed to secure data storage and transfer through decentralized, trustless, peer-to-peer systems with no participation of a third party which records transactions of value using a cryptographic signature [23], [25], [26]. Blockchain started from cryptocurrency, grew in assets and credit field, and increasingly found its place in information and communication field. Various industries have realized the value of the blockchain and how this technology is secure and  reliable as a technical solution. This technical solution allows users to contribute jointly in data computing, storage, authenticity verification and the preserving the reliable database [27], [28].

Early research shows that blockchain technology could potentially provide solutions to some of the challenges faced by power systems and it can be used for different concepts of the power system due to the decentralized structure (e.g. privacy and security, energy trading and marketing, using new communication and smart tools). Regarding privacy and security in power systems, Kanhere et al. [29] applied blockchain to Direct Load Control to protect user privacy and security of communications. Yang et al. [30] proposed an algorithm, applied to a self-organized cyber-physical power system, which has short blockchain construction time and achieves better data block exchange performance. G. Liang [31] showed how blockchain technology can be used to enhance the robustness and security of the power grid.

In terms of energy trading and marketing, recently, blockchain has been an interesting topic for many researchers and companies. Aitzhan and Svetinovic [32] proposed using blockchain to build a decentralized energy trading system. To secure the energy trading transactions in their token-based system, multi-signatures and anonymous encrypted message propagation streams were used. K. Mannaro et al. [33] developed a blockchain-based platform to recommend the best trading strategy for prosumers in the renewable energy market.

In addition, some blockchain projects have focused on energy trading ,especially renewable energy, and smart tools. It should be noted, most of these projects are still in the testing phase or under development. The PWR.Company developed Ethereum-based solutions for trading renewable energy and installed deep cycle batteries for consumers for power storage to stabilize the grid, instead of selling the energy immediately. SolarCoin, PowerLedger, Key2Energy and TheSunExchange aim is to increase solar energy production and facilitate the trade of this type of energy [34]. NRGcoin [35] is currently at the conceptual stage, uses smart contracts framework which is based on Ethereum for trading an energy-based cryptocurrency. Regardless of the retail value of electricity, one NRGcoin is equivalent to one kWh. Share&Charge [36] developed a network of electric vehicle (EV) charging stations and owners of charging stations can register their station and set tariffs for charging. Finally, some projects are focused on smart metering tools for increasing the performance of power systems in fields of energy trading and privacy and security such as Bankymoon [37] and the Electron company [38]. these examples demonstrate that the research activities conducted in this field are significant. However, no bibliometric analysis has been done to report the impacts and trends of such researches.

Bibliometric allows researchers to understand the characteristics, structure, and patterns of research activities. Also, the research activities are combined into a realistic trend of a research domain by this statistical analysis. This involves literature studies of scientific activities in different contexts such as publications, authors, institutions, citations, and countries. Moreover, this method reports on the comprehensive evaluation of the expansion of research fields [39].

The purpose of this study is to well understand the state-of-the-art application of blockchain in power systems. It is vital to identify top-tier researchers, organi-zation and institutes, and collaboration amongst them as well as hot topics. To address these questions, we aim to make a bibliometric analysis on relevant papers published in the Web of Science from 2009 to 2019.

The outline of this paper is as follows. We present the research method in Section 2. Thereafter in Section 3 findings and information about using blockchain in power systems are presented. Section 4 is the conclusion to the study.

## 2. Methodology

The citation analysis in academic papers was initiated by Garfield [40]. Bibliometrics was defined by Pritchard as the application of mathematics and statistical methods to books and other media of communication and is the oldest research methods in library and information science [41]. Bibliometric contains various applications from information science, sociology and history of science to research evaluation [42]. This method is used to evaluate, monitor and visualize the structure of scientific fields [43]. Bibliometric methods can be divided in two parts: general instructions and publication analysis. For general instructions, researchers show how to avoid possible sources of error in the search process by showing how to search article using a search engine. However, the evaluation of publication such as impact factor, citations, publisher, and country described in publication analysis [39]. In general, citation analysis and content analysis are two widely used bibliometric methods. Citation analysis helps to identify core literatures, journals, countries, etc., and shows a relationship between citing and cited works in a research area [44]. For examples M. Dabbagh [45] studied the Evolution of Blockchain, and analyzed scientific production of Geographical Information System (GIS) in Web of Science, and WL.Woon et al. [46] presented a bibliometric based study on distributed generation. Three main bibliometrics data sources for searching the literature are Web of Science, Scopus, Google Scholar. These sources are generally used to rank journals in terms of their productivity and the total citation received to indicate the journal impact, prestige or influence. In this paper, the WoS data is selected to complete the bibliometric analysis based on the following reasons.

Web of science is the most famous tool for bibliometrics analysis and until the creation of Scopus and Google Scholar in 2004 [47], and it used to be the only tool and contains great features. This bibliometrics tool has over 12,000 titles of journals since 1900 to present, covers 45 Languages, and provides citation analysis by author, country, document type, institution, language, publication year, source title, subject area and funding information. Web of Science contains citation maps which helps to visualize the result of the citing references. The cited reference search in WOS is a unique feature that cannot be found in any other databases [48]. In addition, 94% of Scopus highest impact factor journals were indexed in WoS [49].

After using Web of Science as the search engine of this study, we identified some related keywords to start the process of extracting papers. There are some equivalent for Blockchain such as distributed ledger and cryptocurrency [50]. Also, the power network sometimes named by power network or electrical power system [51]. So, the inquiry to collect the data for bibliometric analysis was as follows: (TS = ((Blockchain OR Distributed ledger OR Cryptocurrency) AND (Power System OR Power Network OR Electrical Power Network))). The time period for this study, was limited to the past decade (between 2009 and 2019). As a result, we detected a total of 291 publications from various journals, books, conferences, and patterns. For this analysis, two following databases were selected: "Web of Science Core Collection" and "Current Contents Connect" . To remove unrelated publications such as patterns and non-English publications, we excluded other databases. As a result, 271 articles were secured for this analysis' purposes. This process is shown in Figure 2 . The criteria of this bibliometric analysis are: (a) productivity, (b) research areas, (c) institutions, (d) authors, (e) Impact Publishers,(f) highly cited articles and (g) keyword frequency. Figure 2 is provided for illustration. It should be noted, there is no result before 2014. So, in the rest of this research, presented results will be limited to 2014-2019.

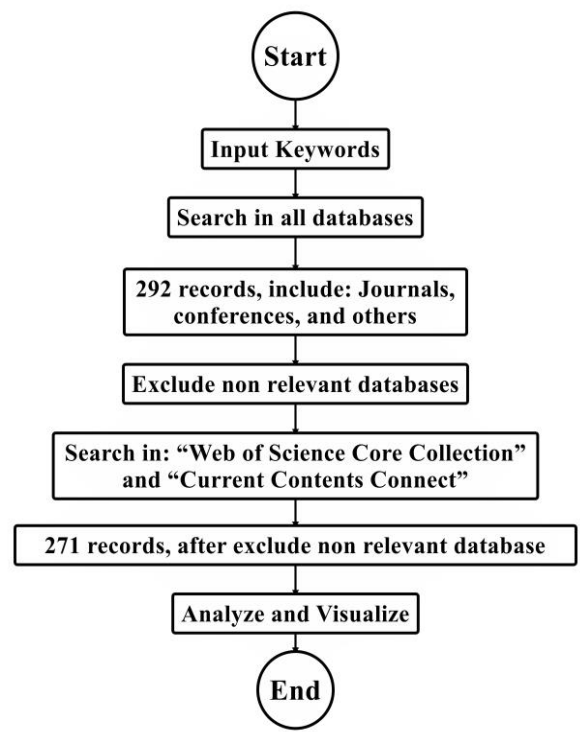

**Fig. 2 The schematic of data collection process**

## 3. Findings

In this section, we discuss the finding of the bibliometric analysis for blockchain application in power systems. The results detect high-quality research to support researchers enhancing research in this field. Finding section is divided into 7 sub-topics: productivity, research areas, institutions, impact journals, authors, highly cited articles and keyword frequency. Fig. 3 shows the number of publications between 2014–2019.

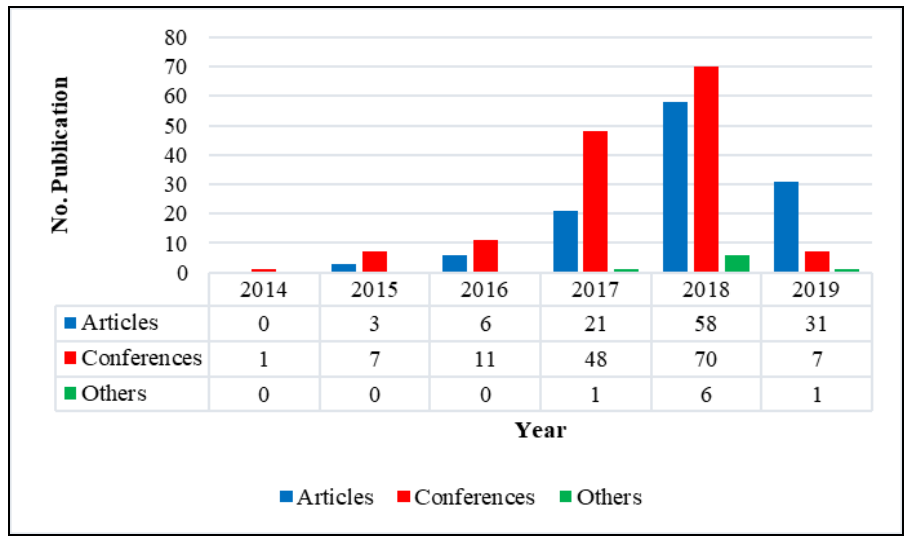

**Fig. 3 The Number of publications**

Figure 3 shows three categories of publications including journals, conferences, and other types (reviews, editorials and abstracts) which are extracted from various studies related to blockchain application in power systems. The conference category has the highest proportion of the total publications by 56.16% in this time period. However, in 2019, as of 21st-Jun-2019, the share of articles has been about 79.5% (30 of 39) which shows a significant increase in this type of documents. This value for 2018 was 43.2%. It is more likely that the journal publications would increase even more in the remaining of 2019.

As it mentioned previously, citation analysis is used to recognize the frequency of the journals and to evaluate researchers' performance. Also, this analysis provides an overview of the topic studied and information about researchers to other researchers using common references. It has been realizing that there are two major types of publications in the academic research study. These publications focusing on originality and developers of the contents to show the significance of research.

The citation is a way of showing the evidence of material in the publications to illustrate the increasing number of research activities that contributed to the high impact of publications. Figure 4 demonstrates the citations received by the publications over the last 6 years. As the number of publications has increased the number of citations has been also increased. The earlier publication which stays in the database for a longer period of time, has the higher chance to be cited. The average number of citations is about 218 annually during 2014 – 2018. The number of annual citations shows a positive trend with a distinct peak occurring in 2018. The citation has increased by about five-fold in 2018 when compared to 2017. Given that the study was conducted in June 2019, it is expected to see the number of citations to be higher than 2018 for 2019.

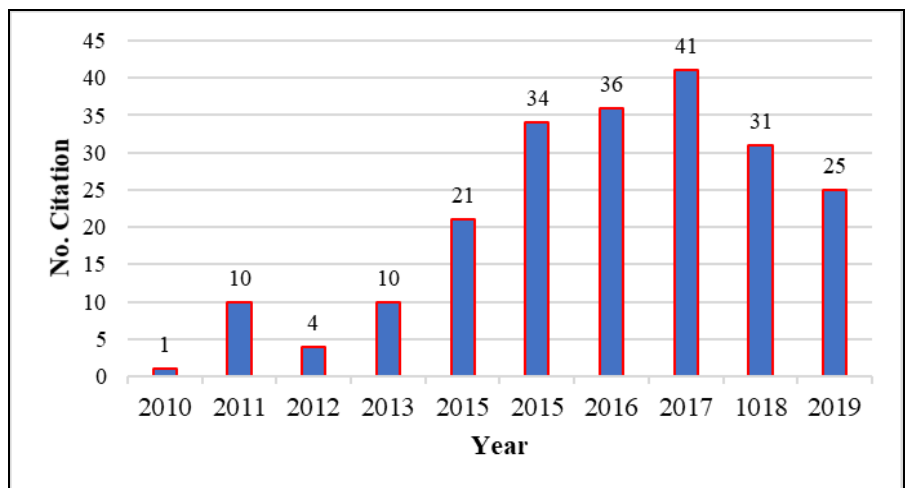

**Fig. 4 The Number of Citation**

### 3.1. Productivity

The productivity of the countries, which refers to the frequency or the number of publications, is presented in this section. A study of productivity growth of articles reflects the attentions and overall strength of different countries in the related research fields. It also shows strengthen of the research mechanisms while leading research involving analysis on blockchain application in power systems. The focus on productivity analysis helps to enhance and improve the production efficiency of the publications. This also assesses which countries have produced more publications.

Figure 5 shows that China and the United States are the lead countries in the number of publications and data show these two countries contributed to almost half of the entire publications related to blockchain application in power systems. As Table 1 shows, this is followed by South Korea, England, and Australia.

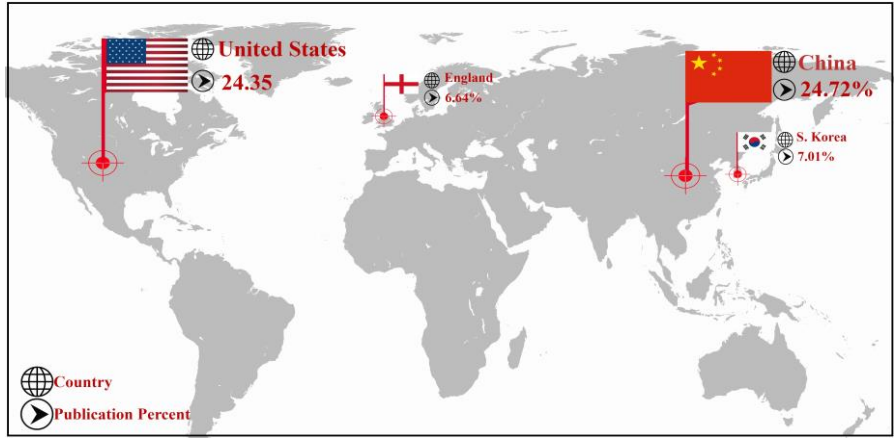

**Fig. 5 The Most Productive Countries**

**Table 1 Productivity**

| Country | Publication (No) | Publication (%) |
| --- | --- | --- |
| China | 67 | 24.72 |
| United States | 66 | 24.35 |
| South Korea | 19 | 7.01 |
| England | 18 | 6.64 |
| Australia | 16 | 5.90 |
| Italy | 15 | 5.54 |
| Singapore | 15 | 5.54 |
| Germany | 14 | 5.16 |
| India | 12 | 4.48 |
| France | 10 | 3.79 |
| Canada | 9 | 3.32 |
| Romania | 9 | 3.32 |
| Japan | 8 | 2.95 |
| Norway | 6 | 2.21 |
| Austria | 5 | 1.84 |
| Poland | 5 | 1.84 |
| Russia | 5 | 1.84 |
| Scotland | 5 | 1.84 |
| Switzerland | 5 | 1.84 |
| United Arab Emirates | 5 | 1.84 |

## 3.2. Research areas

To measure the research performance based on citation and publication rates, researchers use research areas which shows the trend of the publication over time. Research areas develop a logical understanding of explicit research areas and how these challenge other areas in different sectors of the industries. Table 2 shows more details about research areas.

Table 2 shows that the majority of the publications fall under computer science and engineering areas. In this regard, computer science and engineering are the two main research areas for blockchain application in power systems.

**Table 2 Research Areas**

| Research Areas | Publication (No) | Publication (%) |
|---|---|---|
| Computer Science | 187 | 69.00 |
| Engineering | 154 | 56.82 |
| Energy Fuels | 97 | 35.79 |
| Telecommunication | 64 | 23.61 |
| Business Economics | 60 | 22.14 |
| Communication | 47 | 17.34 |
| Mathematics | 26 | 9.59 |
| Automation Control Systems | 25 | 9.22 |
| Science Technology | 17 | 6.27 |
| Instrument Instrumentation | 10 | 3.69 |
| Government Law | 4 | 1.47 |
| Physics | 4 | 1.47 |
| Educational Research | 3 | 1.10 |
| Materials Science | 3 | 1.10 |
| Robotics | 3 | 1.10 |
| Transportation | 3 | 1.10 |
| Optics | 2 | 0.74 |
| Social Work | 2 | 0.74 |
| Development Studies | 1 | 0.37 |
| Mechanics | 1 | 0.37 |
| Nuclear Science Technology | 1 | 0.37 |

### 3.3. Institutions

This section discusses the number of publications noted according to various institutions and measures different institution's quality according to their publications. Also, it identifies which of these institutions are currently active. Table 3 lists the institutions which conducted research related to blockchain application in power systems. The table shows that institutions in China in total have the highest number of publications. According to this table, the Chinese Academy of Sciences and Nanyang Technological University have the highest number of publications. The most prominent institutions in Asia are located in China. It seems that the speed of publication about using blockchain in power systems, in China is much faster than the other countries in the world. This evidence suggests that there is keen competition among institutions across China in terms publications.

**Table 3 Institution**

| Institution | Publication (No) | Publication (%) | Country |
|---|---|---|---|
| Chinese Academy of Sciences | 10 | 10 | China |
| Nanyang Technological University | 10 | 10 | Singapore |
| National University of Singapore | 6 | 6 | Singapore |
| University of California | 6 | 6 | United States |
| Beijing University of P&T | 5 | 5 | China |
| Politehnica University of Bucharest | 5 | 5 | Romania |
| Shanghai Jiao Tong University | 5 | 5 | China |
| UESTC | 5 | 5 | China |
| UNSW Sydney | 5 | 5 | Australia |
| University of Aalborg | 4 | 4 | Denmark |
| Chung-Ang University | 4 | 4 | South Korea |
| NUDT | 4 | 4 | China |
| Tsinghua University | 4 | 4 | China |
| United States Department of Energy | 4 | 4 | United States |
| University of Illinois | 4 | 4 | United States |
| The University of Newcastle | 4 | 4 | Australia |
| Chongqing University | 3 | 3 | China |

### 3.4. Authors

This section discusses the number of publications noted according to authors in various countries to identify who is the most active in terms of authorship. Table 4 lists the findings for authors who are the most productive. As the table illustrates, the majority of the authors are from China, Australia, Denmark, Italy, and Singapore. It appears some other countries such as Greece, Canada, Norway, and Japan are also able to contribute to many publications.

**Table 4 List of authors**

| Authors | Publication (No) | Publication (%) | Country |
|---|---|---|---|
| Yan Chen | 4 | 1.47% | China |
| Pietro Danzi | 4 | 1.47% | Denmark |
| Aggelos Kiayias | 4 | 1.47% | Greece |
| Petar Popovski | 4 | 1.47% | Denmark |
| Prateek Saxena | 4 | 1.47% | Singapore |
| Cedomir Stefanovic | 4 | 1.47% | Denmark |
| Jun Wang | 4 | 1.47% | Australia |
| Xiaonan Wang | 4 | 1.47% | Singapore |
| Yong Yuan | 4 | 1.47% | China |
| Yan Zhang | 4 | 1.47% | Norway |
| Ryosuke Abe | 3 | 1.11% | Japan |
| Matthew Davison | 3 | 1.11% | Australia |
| Maria Di Silvestre | 3 | 1.11% | Italy |
| Fei Yue Wang | 3 | 1.11% | China |
| Nikos Leonardos | 3 | 1.11% | Greece |
| Yang Li | 3 | 1.11% | Canada |
| Loi Luu | 3 | 1.11% | Singapore |
| Cristina Roscia | 3 | 1.11% | Italy |
| Eleonora Sanseverino | 3 | 1.11% | Italy |
| Terrence Summers | 3 | 1.11% | Australia |
| Christopher Townsend | 3 | 1.11% | Australia |
| Ping Wang | 3 | 1.11% | Canada |
| Hui Yang | 3 | 1.11% | China |
| Xiaosong Zhang | 3 | 1.11% | China |
| Gaetano Zizzo | 3 | 1.11% | Italy |

## 3.5. Publishers

This section discusses the list of publishers which published the most publications about blockchain application in power systems. This section is important as it shows the most leading journals, conferences, and book series in publications and the ones which have the highest citations. This information helps researchers to identify the high-quality journals and conferences to strengthen their work by publishing in them. Table 5 lists some publishers titles with the greatest number of publications in the field. It shows that the greatest number of publications belongs to the IEEE Access journal and Lecture Notes in Computer Sciences book series and followed by other journals such as Energies and Sensors.

Table 5 demonstrates that IEEE Journals and Conferences are major publishers in terms of blockchain application in power systems. It shows that Lecture Notes in Computer Science received 67051 citations over the years followed by Applied Energy and Sensors with 42891 and 25150 citations respectively. This table also illustrates that journals with dominant citations per document by a remarkable difference from the rest, are IEEE Internet of Things Journal, Applied Energy, and IEEE Transactions on Industrial Informatics. As a whole, the quality of high impact journals attracts researchers to publish their articles because it widely read by the other researcher and increases their citations.

**Table 5 List of Publishers**

| Title | Type | P | TC | CD | CPD |
|---|---|---|---|---|---|
| IEEE Access | Journals | 12 | 19132 | 3277 | 4.944 |
| Lecture Notes in Computer Science | Book Series | 12 | 67051 | 25610 | 1.120 |
| Energies | Journals | 8 | 12160 | 3202 | 3.178 |
| Sensors | Journals | 6 | 25150 | 5692 | 3.715 |
| ICRERA | Conferences | 5 | 251 | 99 | 1.873 |
| Applied Energy | Journals | 4 | 42891 | 4416 | 9.593 |
| IEEE ICSGC | Conferences | 3 | 210 | 90 | 1.680 |
| Future Generation Computer Systems | Journals | 3 | 4600 | 658 | 6.897 |
| IEEE Internet of Things Journal | Journals | 3 | 4529 | 369 | 11.613 |
| IEEE Spectrum | Journals | 3 | 383 | 133 | 0.934 |
| IEEE Transactions on Industrial Informatics | Journals | 3 | 6348 | 676 | 8.878 |
| ICPADS | Conferences | 3 | 256 | 134 | 0.959 |
| Sustainability | Journals | 3 | 13827 | 3781 | 3.029 |

P: Publication No; TC: Total Cites; CD: Cited documents; CPD: Citations per document (2015-2018)

## 3.6. Highly-cited articles

This section illustrates the quality and influence of research done in using blockchain in power systems by assessing the number of citations received by each publication. Table 6 lists the top 15 most cited publications, number of times cited, type to about 5.53% of the total publications. Moreover, the top highly-cited publication was published 4 years ago, showing compliance with the concept that the longer the publications have been in the database, the higher the number of citations accumulated. Even though the blockchain is a new advent technology and there is no publication about blockchain application in power systems prior to 2014, the number of citations for publications in this field is high.

Of the articles published , the most cited was "Blockchain technology in the chemical industry: Machine-to-machine electricity market". This paper investigated blockchain application and presented a scenario including two electricity producers and one electricity consumer trading with together based on blockchain. It can be concluded that highly cited articles are high quality research in which the researcher recognizes other author's findings, ideas, methods, and influence in certain fields. As a whole, if the topic in articles is more interesting, it increases journal citations particularly when the subject is more special.

**Table 6 Top 15 Highly-Cited Publications**

| Title | Times Cited |
|---|---|
| The Bitcoin Backbone Protocol: Analysis and Applications | 116 |
| Blockchain technology in the chemical industry: Machine-to-machine electricity market | 63 |
| Security and Privacy in Decentralized Energy Trading Through Multi-Signatures, Blockchain and Anonymous Messaging Streams | 61 |
| Enabling Localized Peer-to-Peer Electricity Trading Among Plug-in Hybrid Electric Vehicles Using Consortium Blockchains | 60 |
| Blockstack: A Global Naming and Storage System Secured by Blockchains | 52 |
| A Secure Sharding Protocol for Open Blockchains | 51 |
| Industry 4.0: state of the art and future trends | 45 |
| A blockchain-based smart grid: towards sustainable local energy markets | 41 |
| Blockchain Based Decentralized Management of Demand Response Programs in Smart Energy Grids | 38 |
| Analyzing the Bitcoin Network: The First Four Years | 26 |
| Consortium Blockchain for Secure Energy Trading in Industrial Internet of Things | 25 |
| Research on the Technology and Economic Calculation Model of Power Transmission Line Considering Environmental Benefits | 22 |
| Citizen utilities: The emerging power paradigm | 21 |
| Privacy-Preserving and Efficient Aggregation Based on Blockchain for Power Grid Communications in Smart Communities | 20 |
| Cryptocurrencies Without Proof of Work | 20 |

414

## 3.7. Keywords frequency

This section discusses the type of keywords which are frequently used by researchers. These keywords could be used to analyze and identify research trends and gaps. Table 7 provides a list of unique keywords and title occurrences. This list was derived from a total of 5,958 keywords and 701 titles that had appeared in 271 publications between 2014 and 2019.

Table 7 shows that the most relevant title and keyword is blockchain. This table also shows that blockchain and power system are consistently used in the literature.

**Table 7 Frequency of Keywords in Titles and Abstracts**

| Titles | Frequency | Keywords | Frequency |
|---|---|---|---|
| Blockchain | 93 | Blockchain | 364 |
| System | 33 | Power system | 136 |
| Bitcoin | 21 | Protocol | 95 |
| Analysis | 12 | Energy | 90 |
| Microgrid | 12 | Market | 84 |
| Application | 11 | Miner | 75 |
| Attack | 10 | Block | 58 |
| Power system | 10 | Cost | 55 |
| Cryptocurrency | 9 | Smart contract | 46 |
| Mining | 8 | Vehicle | 46 |
| Security | 8 | Mining | 43 |
| IOT | 7 | Management | 40 |
| Smart grid | 6 | Privacy | 40 |
| Energy trading | 6 | Method | 38 |
| Privacy | 5 | Grid | 36 |

To provide an in-depth analysis, Figure 6 presents a word map based on a content analysis of the publications. According to this map, the keywords are divided into 3 clusters in which two clusters are more specific. One of these clusters highlighted by key terms which are related to "cryptocurrency", "bitcoin", "protocol", "attack", "reward" and another one contains keywords such as "microgrid", "energy", "market", "electric vehicle". In addition, ''power system'', "blockchain", and "energy" were noted as terms that act as links between the research topics.

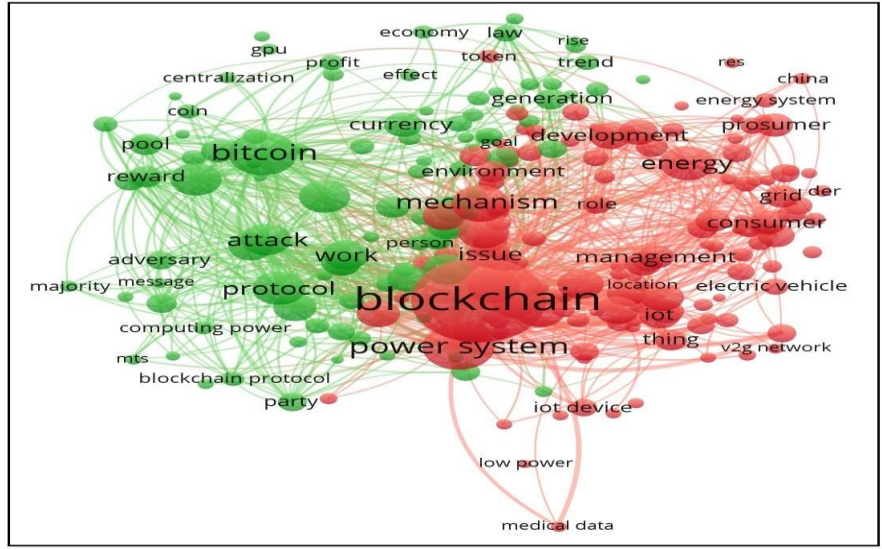

**Fig. 6  Keywords Map**

## 4. Conclusions

In this paper, we used WoS as the literature source for the bibliometric analysis of blockchain application in power system from January 2014 to June 21, 2019. Seven criteria including productivity, research areas, institutions, authors, impact publishers, highly cited articles, and keyword frequency have been used in this study. Using these criteria, we uncovered the global trends and frontiers related to our subject. Between 2014 and 2018, it was noted that the number of publications related to blockchain application in power system had increased with an average annual growth rate of 418%. The analysis also indicated that the trend of blockchain publications experienced speedy progress with increased article publications and citations during this time period.

It was noted that China and the United States are the lead countries with the most publications produced in academic research. Then, we showed that the majority of the publications fall under computer science and engineering.

Our analysis had indicated that IEEE Journals and Conferences are the major publishers in terms of blockchain application in power systems. This study also highlighted the active authors in terms of publications in different countries. Among the top 15 most active authors, there are 10 authors with 4 publications in this field. Finally, a map analysis of keyword frequencies had been used to describe the trends and research directions for future studies in blockchain application in power system field.

# References

[1] L. Abdallah and T. El-Shennawy, "Reducing Carbon Dioxide Emissions from Electricity Sector Using Smart Electric Grid Applications," *J. Eng.*, vol. 2013, pp. 1–8, Apr. 2013.

[2] H. M. Ruzbahani and H. Karimipour, "Optimal incentive-based demand response management of smart households," in *2018 IEEE/IAS 54th Industrial and Commercial Power Systems Technical Conference (I&CPS)*, 2018, pp. 1–7.

[3] H. Karimipour, A. Dehghantanha, R. M. Parizi, K.-K. R. Choo, and H. Leung, "A Deep and Scalable Unsupervised Machine Learning System for Cyber-Attack Detection in Large-Scale Smart Grids," *IEEE Access*, vol. 7, pp. 80778–80788, 2019.

[4] D. Tan, "Energy Challenge, Power Electronics & Systems (PEAS) Technology and Grid Modernization," *CPSS Trans. Power Electron. Appl.*, vol. 2, no. 1, pp. 3–11, Apr. 2017.

[5] H. Rouzbahani, … A. R. preprint arXiv, and undefined 2019, "Smart Households Demand Response Management with Micro Grid," *arxiv.org*.

[6] H. Karimipour, V. D.-I. transactions on smart grid, and undefined 2015, "Extended Kalman filter-based parallel dynamic state estimation," *ieeexplore.ieee.org*.

[7] F. Ghalavand *et al.*, "Microgrid Islanding Detection Based on Mathematical Morphology," *Energies*, vol. 11, no. 10, p. 2696, Oct. 2018.

[8] J. Sakhnini, H. Karimipour, and A. Dehghantanha, "Smart Grid Cyber Attacks Detection using Supervised Learning and Heuristic Feature Selection," Jul. 2019.

[9] H. Karimipour, V. D.-2014 N. A. Power, and undefined 2014, "On detailed synchronous generator modeling for massively parallel dynamic state estimation," *ieeexplore.ieee.org*.

[10] H. Karimipour and V. Dinavahi, "Robust Massively Parallel Dynamic State Estimation of Power Systems Against Cyber-Attack," *IEEE Access*, vol. 6, pp. 2984–2995, 2018.

[11] H. Karimipour, V. D.-2013 N. A. Power, and undefined 2013, "Accelerated parallel WLS state estimation for large-scale power systems on GPU," *ieeexplore.ieee.org*.

[12] H. Karimipour, V. D.-2017 I. International, and undefined 2017, "On false data injection attack against dynamic state estimation on smart power grids," *ieeexplore.ieee.org*.

[13] H. Karimipour and V. Dinavahi, "On false data injection attack against dynamic state estimation on smart power grids," in *2017 IEEE International Conference on Smart Energy Grid Engineering (SEGE)*, 2017, pp. 388–393.

[14] V. Dinavahi and H. Karimipour, "Parallel relaxation-based joint dynamic

state estimation of large-scale power systems," *IET Gener. Transm. Distrib.*, vol. 10, no. 2, pp. 452–459, Feb. 2016.

[15] F. Ghalavand, I. Al-Omari, … H. K.-… C. on S., and undefined 2018, "Hybrid islanding detection for ac/dc network using DC-link voltage," *ieeexplore.ieee.org*.

[16] S. Zhou and M. A. Brown, "Smart meter deployment in Europe: A comparative case study on the impacts of national policy schemes," *J. Clean. Prod.*, vol. 144, pp. 22–32, Feb. 2017.

[17] A. J. Goldsmith and S. B. Wicker, "Design challenges for energy-constrained ad hoc wireless networks," *IEEE Wirel. Commun.*, vol. 9, no. 4, pp. 8–27, Aug. 2002.

[18] R. Mijumbi, J. Serrat, J.-L. Gorricho, N. Bouten, F. De Turck, and R. Boutaba, "Network Function Virtualization: State-of-the-Art and Research Challenges," *IEEE Commun. Surv. Tutorials*, vol. 18, no. 1, pp. 236–262, 2016.

[19] F. Ghalavand, B. Alizade, H. Gaber, and H. Karimipour, "Microgrid Islanding Detection Based on Mathematical Morphology," *Energies*, vol. 11, no. 10, p. 2696, Oct. 2018.

[20] U. Ahsan and A. Bais, "Distributed big data management in smart grid," in *2017 26th Wireless and Optical Communication Conference (WOCC)*, 2017, pp. 1–6.

[21] S. Mohammadi, … H. M.-J. of information, and undefined 2019, "Cyber intrusion detection by combined feature selection algorithm," *Elsevier*.

[22] E. Dovom, A. Azmoodeh, … A. D.-J. of S., and undefined 2019, "Fuzzy pattern tree for edge malware detection and categorization in IoT," *Elsevier*.

[23] R. M. Parizi and A. Dehghantanha, "On the Understanding of Gamification in Blockchain Systems," in *2018 6th International Conference on Future Internet of Things and Cloud Workshops (FiCloudW)*, 2018, pp. 214–219.

[24] R. M. Parizi, Amritraj, and A. Dehghantanha, "Smart Contract Programming Languages on Blockchains: An Empirical Evaluation of Usability and Security," Springer, Cham, 2018, pp. 75–91.

[25] P. J. Taylor, T. Dargahi, A. Dehghantanha, R. M. Parizi, and K.-K. R. Choo, "A systematic literature review of blockchain cyber security," *Digit. Commun. Networks*, Feb. 2019.

[26] S. Aggarwal, R. Chaudhary, G. S. Aujla, A. Jindal, A. Dua, and N. Kumar, "EnergyChain," in *Proceedings of the 1st ACM MobiHoc Workshop on Networking and Cybersecurity for Smart Cities - SmartCitiesSecurity'18*, 2018, pp. 1–6.

[27] R. M. Parizi, S. Homayoun, A. Yazdinejad, A. Dehghantanha, and K.-K. R. Choo, "Integrating Privacy Enhancing Techniques into Blockchains Using Sidechains," Jun. 2019.

[28] F. Dai, Y. Shi, N. Meng, L. Wei, and Z. Ye, "From Bitcoin to

cybersecurity: A comparative study of blockchain application and security issues," in *2017 4th International Conference on Systems and Informatics (ICSAI)*, 2017, pp. 975–979.

[29] A. Dorri, F. Luo, S. S. Kanhere, R. Jurdak, and Z. Y. Dong, "A Secure and Efficient Direct Power Load Control Framework Based on Blockchain," Dec. 2018.

[30] T. Yang, F. Zhai, J. Liu, M. Wang, and H. Pen, "Self-organized cyber physical power system blockchain architecture and protocol," *Int. J. Distrib. Sens. Networks*, vol. 14, no. 10, p. 155014771880331, Oct. 2018.

[31] G. Liang, S. R. Weller, F. Luo, J. Zhao, and Z. Y. Dong, "Distributed Blockchain-Based Data Protection Framework for Modern Power Systems Against Cyber Attacks," *IEEE Trans. Smart Grid*, vol. 10, no. 3, pp. 3162–3173, May 2019.

[32] N. Z. Aitzhan and D. Svetinovic, "Security and Privacy in Decentralized Energy Trading Through Multi-Signatures, Blockchain and Anonymous Messaging Streams," *IEEE Trans. Dependable Secur. Comput.*, vol. 15, no. 5, pp. 840–852, Sep. 2018.

[33] K. Mannaro, A. Pinna, and M. Marchesi, "Crypto-trading: Blockchain-oriented energy market," in *2017 AEIT International Annual Conference*, 2017, pp. 1–5.

[34] M. Meisel, S. Wilker, A. G. Goranovi´c, L. Fotiadis, A. Treytl, and T. Sauter, "Blockchain applications in microgrids an overview of current projects and concepts Thilo Sauter TU Wien Blockchain Applications In Microgrids An overview of current projects and concepts."

[35] M. Mihaylov, I. Razo-Zapata, and A. Nowé, "NRGcoin—A Blockchain-based Reward Mechanism for Both Production and Consumption of Renewable Energy," *Transform. Clim. Financ. Green Invest. with Blockchains*, pp. 111–131, Jan. 2018.

[36] F. Vanrykel, D. Ernst, and M. Bourgeois, "Fostering Share&Charge through proper regulation," *Compet. Regul. Netw. Ind.*, vol. 19, no. 1–2, pp. 25–52, Mar. 2018.

[37] R. Chitchyan and J. Murkin, "Review of Blockchain Technology and its Expectations: Case of the Energy Sector," Mar. 2018.

[38] A. V. Vladimirova, "Blockchain Revolution in Global Environmental Governance: Too Good to Be True?," Springer, Cham, 2019, pp. 212–223.

[39] M. F. A. Razak, N. B. Anuar, R. Salleh, and A. Firdaus, "The rise of 'malware': Bibliometric analysis of malware study," *J. Netw. Comput. Appl.*, vol. 75, pp. 58–76, Nov. 2016.

[40] H. D. White, "Pennants for Garfield: bibliometrics and document retrieval," *Scientometrics*, vol. 114, no. 2, pp. 757–778, Feb. 2018.

[41] V. W.-E. B. L. and Information and undefined 2013, "Formalized curiosity: reflecting on the librarian practitioner-researcher," *ejournals.library.ualberta.ca*.

[42] P. Mongeon and A. Paul-Hus, "The journal coverage of Web of Science

and Scopus: a comparative analysis," *Scientometrics*, vol. 106, no. 1, pp. 213–228, Jan. 2016.

[43] J. Koskinen *et al.*, "How to use bibliometric methods in evaluation of scientific research? An example from Finnish schizophrenia research," *Nord. J. Psychiatry*, vol. 62, no. 2, pp. 136–143, Jan. 2008.

[44] A. Pilkington and J. Meredith, "The evolution of the intellectual structure of operations management—1980–2006: A citation/co-citation analysis," *J. Oper. Manag.*, vol. 27, no. 3, pp. 185–202, Jun. 2009.

[45] M. Dabbagh, M. Sookhak, and N. S. Safa, "The Evolution of Blockchain: A Bibliometric Study," *IEEE Access*, vol. 7, pp. 19212–19221, 2019.

[46] W. Woon, H. Zeineldin, S. M.-T. F. and Social, and undefined 2011, "Bibliometric analysis of distributed generation," *Elsevier*.

[47] L. S. Adriaanse and C. Rensleigh, "Comparing Web of Science, Scopus and Google Scholar from an Environmental Sciences perspective," *South African J. Libr. Inf. Sci.*, vol. 77, no. 2, Jan. 2011.

[48] J. Li, J. F. Burnham, T. Lemley, and R. M. Britton, "Citation Analysis: Comparison of Web of Science®, Scopus™, SciFinder®, and Google Scholar," *J. Electron. Resour. Med. Libr.*, vol. 7, no. 3, pp. 196–217, Aug. 2010.

[49] C. López-Illescas, F. de Moya-Anegón, and H. F. Moed, "Coverage and citation impact of oncological journals in the Web of Science and Scopus," *J. Informetr.*, vol. 2, no. 4, pp. 304–316, Oct. 2008.

[50] S. Miau and J.-M. Yang, "Bibliometrics-based evaluation of the Blockchain research trend: 2008 – March 2017," *Technol. Anal. Strateg. Manag.*, vol. 30, no. 9, pp. 1029–1045, Sep. 2018.

[51] E. Hache and A. Palle, "Renewable energy source integration into power networks, research trends and policy implications: A bibliometric and research actors survey analysis," *Energy Policy*, vol. 124, pp. 23–35, Jan. 2019.